\documentclass{aa}
\usepackage{natbib} 
\usepackage[english]{babel}
\usepackage{latexsym}
\usepackage[intlimits,sumlimits]{amsmath}
\usepackage{graphicx}

\newcommand{\Ne}{n_{\rm e}}
\newcommand{\Te}{T_{\rm e}}
\newcommand{\me}{m_{\rm e}}
\newcommand{\sigT}{\sigma_{\rm T}}

\newcommand{\beal}{\begin{align}}
\newcommand{\bsub}{\begin{subequations}}
\newcommand{\esub}{\end{subequations}}
\newcommand{\bmulti}{\begin{multline}}   
\newcommand{\Abst}[1]{\,#1}
\newcommand{\id}{{\,\rm d}}

\newcommand{\pAb}[2]{\frac{\displaystyle\pd #1}{\displaystyle\pd #2}}
\newcommand{\pd}{\partial}
\newcommand{\Abl}[2]{\frac{{\rm d} #1}{{\rm d} #2}}
\newcommand{\vek}[1]{\mbox{\boldmath${#1}$\unboldmath}}

\newcommand{\gkin}{{\mathcal{C}_{\rm k}}}

\newcommand{\gth}{{\mathcal{C}_{\rm th}}}

\newcommand{\ykinp}{{y_{\rm kin}}}

\newcommand{\DFSZ}{{\Delta F_{\,\rm SZ}}}
\newcommand{\DFSZp}{{\Delta F'_{\,\rm SZ}}}

\newcommand{\DISZ}{{I_{\,\rm SZ}}}
\newcommand{\DISZp}{{I'_{\,\rm SZ}}}
\newcommand{\DIth}{{I_{\,\rm th}}}
\newcommand{\DIk}{{I_{\,\rm k}}}
\newcommand{\DIthp}{{\Delta I'_{\rm D, th}}}
\newcommand{\DIkp}{{\Delta I'_{\rm D, k}}}

\newcommand{\betao}{{\beta_{\rm o}}}
\newcommand{\betaopar}{{\beta_{\rm o,\parallel}}}
\newcommand{\bbetao}{{\beta_{\rm o}^2}}

\newcommand{\betac}{{\beta_{\rm c}}}
\newcommand{\betacpar}{{\beta_{\rm c,\parallel}}}

\newcommand{\lesssim}{\mathrel{\hbox{\rlap{\hbox{\lower4pt\hbox{$\sim$}}}\hbox{$<$}}}}
\newcommand{\gtrsim}{\mathrel{\hbox{\rlap{\hbox{\lower4pt\hbox{$\sim$}}}\hbox{$>$}}}} 

\newcommand{\x}{{\hat{x}}}
\newcommand{\xn}{x}

\newcommand{\kB}{k\,}

\newcommand{\thr}{{\theta_{\rm r}}}
\newcommand{\thrp}{{\theta'_{\rm r}}}

\newcommand{\murp}{{\mu'_{\rm r}}}

\newcommand{\plotwd}{8.2cm}
\newcommand{\plotwdtwo}{16.4cm}

\begin{document}

\titlerunning{Clusters of galaxies in the microwave band} 
\title{Clusters of galaxies in the microwave band: influence \\of the motion of the Solar System}

\author{J. Chluba\inst{1}, G. H{\"u}tsi\inst{1} \and R.A. Sunyaev\inst{1,2}}

\institute{Max-Planck-Institut f\"ur Astrophysik, Karl-Schwarzschild-Str. 1,
86740 Garching bei M\"unchen, Germany
\and Space Research Institute, Russian Academy of Sciences, Profsoyuznaya 84/32 Moscow, Russia}

\offprints{J. Chluba, \\ \email{jchluba@mpa-garching.mpg.de}}
\date{Received / Accepted}

\abstract {In this work we consider the changes of the SZ cluster brightness,
flux and number counts induced by the motion of the Solar System with respect
to the frame defined by the cosmic microwave background (CMB). These changes
are connected with the Doppler effect and aberration and exhibit a strong
spectral and spatial dependence.
The correction to the SZ cluster brightness and flux has an amplitude and
spectral dependence, which is similar to the first order cluster peculiar
velocity correction to the thermal SZ effect.
Due to the change in the received cluster CMB flux the motion of the Solar
System induces a dipolar asymmetry in the observed number of clusters above a
given flux level. Similar effects were discussed for $\gamma$-ray bursts and
radio galaxies, but here, due to the very peculiar frequency-dependence of the
thermal SZ effect, the number of observed clusters in one direction of the sky
can be both, decreased or increased depending on the frequency band. A
detection of this asymmetry should be possible using future full sky CMB
experiments with mJy sensitivities. 
\keywords{Galaxies: clusters --- Cosmology: cosmic microwave background,
spectral distortions, observations}}

\maketitle

\section{Introduction}
Due to the thermal SZ effect \citep{Suny72, Sun80a} clusters of galaxies
(after our own Galaxy) are one the most important and brightest foreground
sources for CMB experiments devoted to studying the primordial temperature
anisotropies.
Given the strong and very peculiar frequency-dependence of the SZ signature
(the flux changes sign at $\nu\sim 217\,$GHz) it is possible to extract these
sources and thereby open the way for deeper investigations of the primordial
temperature anisotropies, which for $l\gtrsim 3000$ are weaker than the
fluctuations due to clusters of galaxies.
Within the next 5 years several CMB experiments like {\sc Planck}, {\sc Spt},
{\sc Act}, {\sc Quiet}, {\sc Apex} and {\sc Ami} will perform deep searches
for clusters with sensitivity limits at the level of $1-10\,$mJy and in the
future CMB missions such as {\sc Cmbpol} should reach sensitivities 20-100
times better than those of {\sc Planck} by using even currently existing
technology \citep{Church2000}. Many tens of thousands of clusters will be
detected allowing to carry out detailed studies of cluster physics and to
place constraints on parameters of the Universe like the Hubble parameter, the
baryonic matter, dark matter and dark energy content \citep[for review
see][]{Birkinshaw99, Carlstrom2002}.

Under this perspective several groups have derived relativistic corrections to
the thermal \citep{Suny72} and kinetic SZ \citep{Sun80} effect
as series expansions in the dimensionless electron temperature, $\kB\Te/\me
c^2$, and the clusters peculiar velocity, $\betac=v_{\rm c}/c$
\citep{Rephaeli1995, Chall98, Itoh98a, Itoh98b, Sazonov98}. 

Motivated by the rapid developments in CMB technology the purpose of this
paper is to take into account the changes in the SZ signal, which are induced
by the motion of the Solar System relative to the CMB rest frame. Assuming
that the observed CMB dipole is fully motion-induced it implies that the Solar
System is moving with a velocity of $\betao=v_{\rm o}/c=1.241\cdot 10^{-3}$
towards the direction $(l, b)=(264.14^\circ\pm 0.15^\circ, 48.26^\circ\pm
0.15^\circ)$ \citep{Smoot77,Fixsen96}.
As will be shown here, in the lowest order of $\betao$
the motion-induced correction to the thermal SZ effect (th-SZ) exhibits an
amplitude and spectral dependence, which is similar to the first order
$\betac$ correction to the th-SZ, i.e. the SZ signal $\propto
\tau\,\betac\,\kB\Te/\me c^2$, with Thomson optical depth $\tau$,
whereas the observers frame transformation of the kinetic SZ effect (k-SZ)
leads to a much smaller y-type spectral distortion with effective
$y$-parameter $\propto \tau\,\betac\,\betao$.

Future CMB experiments like {\sc Planck}, {\sc Spt} and {\sc Act} will not
resolve the central regions for most of the detected clusters.
Therefore here we are not only discussing the change in the {\it brightness}
of the CMB in the direction of a cluster but also the corrections to the {\it
flux} as measured for unresolved clusters due to both the motion-induced
change of surface brightness and the apparent change of their {\it angular
dimension}. All these changes are connected with the Doppler effect and
aberration, which also influence the primordial temperature fluctuations
as discussed by \citet{Chall2002}.

Another important consequence of the motion of the Solar System with respect
to the CMB rest frame is the dipole anisotropy induced in the deep {\it number
counts} of sources. This effect was discussed earlier in connection with the
distribution of $\gamma$-ray bursts \citep{Maoz94,Scharf95} -- identical to
the Compton-Getting effect \citep{Compton35} for cosmic rays -- and radio and
IR sources \citep{Ellis84, Baleisis1998, Blake2002}.
The motion-induced change in the source number counts strongly depends on the
slope of $\log N$-$\log F$ curve and the spectral index of the source
\citep{Ellis84}, which makes it possible to distinguish the signals arising
from different astrophysical populations.
Here we show that a similar effect arises for the number counts of SZ
clusters. Due to the very peculiar frequency-dependence of the th-SZ, the
number of observed clusters in one direction of the sky can be both, decreased
or increased depending on the frequency band.

\section{General transformation laws}
\label{sec:Trans}
A photon of frequency $\nu$ propagating along the direction
$\vek{n}=(\phi,\theta)$ in the CMB rest frame $S$ due to {\it Doppler
boosting} and {\it aberration} is received at a frequency $\nu'$ in the
direction $\vek{n}'=(\phi,\theta')$ by an observer which is moving with the
velocity $\betao=v_{\rm o}/c$ along the $z$-axis:
\beal
\label{eq:trans} 
\nu&=\gamma\nu'(1-\betao\mu')
&\mu&=\frac{\mu'-\betao}{1-\betao\,\mu'}
\Abst{.}
\end{align}
Here $\gamma=1/\sqrt{1-\bbetao}$ is the Lorentz factor, $\mu=\cos \theta$ and
all the primed quantities\footnote{In the following prime denotes that the
corresponding quantity is given in the rest frame of the moving observer.}
denote the corresponding variables in the observers frame $S'$. It was also
assumed that the $z'$-axis is aligned with the direction of the motion.
For a given spatial and spectral distribution of photons in $S$, in lowest
order of $\betao$ the Doppler effect leads to spectral distortions, whereas
due to aberration the signal on the sky is only redistributed.

\subsection*{Transformation of the spectral intensity}
The transformation of the spectral intensity (or equivalently the surface
brightness) $I(\nu, \vec{n})$ at frequency $\nu$ and in the direction
$\vec{n}$ on the sky into the frame $S'$ can be performed using the invariance
properties of the occupation number, $n(\nu, \vec{n})=I(\nu, \vec{n})/\nu^3$:
\beal
\label{eq:transI} 
&I'(\nu', \vec{n}')=\frac{{\nu'}^3}{\nu^3}\,I(\nu, \vec{n})
\Abst{.}
\end{align}
Here $I'(\nu', \vec{n}')$ is the spectral intensity at frequency $\nu'$ in the
direction $\vec{n}'$ as given in the rest frame of the observer. 
In lowest order of $\betao$ it is possible to separate the effects of Doppler
boosting and aberration:
\bsub
\bmulti
\label{eq:transIapp0} 
I'(\nu', \vec{n}')\approx I(\nu',\vec{n}')+\Delta I_{\rm D}(\nu',\vec{n}')+\Delta I_{\rm A}(\nu',\vec{n}')
\Abst{.}
\end{multline}
with the Doppler and aberration correction
\label{eq:transIapp} 
\beal
\Delta I'_{\rm D}(\nu',\vec{n}')
\label{eq:transIappa} 
&\approx \betao\,\mu'\left[3-\nu'\partial_{\nu'}\right]I(\nu',\vec{n}')
\\[1.5mm]
\label{eq:transIappb} 
\Delta I'_{\rm A}(\nu',\vec{n}')&\approx\betao\,\sqrt{1-\mu'^2}\,\partial_{\theta'}I(\nu', \vec{n}')
\Abst{.}
\end{align}
\esub
Equation \eqref{eq:transIappa} only includes the effects due to Doppler
boosting, whereas \eqref{eq:transIappb} arises solely due to aberration.

With \eqref{eq:transIapp} it becomes clear that in first order of $\betao$ any
maximum or minimum of the intensity distribution on the sky will suffer only
from Doppler boosting. This implies that due to aberration the positions of
the central regions of clusters of galaxies will only be redistributed on the
sky: in the direction of the motion clusters will appear to be closer to each
other while in the opposite direction their angular separation will seem to be
bigger.
Another consequence of the observers motion is that a cluster with angular
extension $\Delta\ll 1$ in $S$, will appear to have a size
$\Delta'=\Delta[1-\betaopar]$ in the observers frame $S'$.
Therefore in $S'$ a clusters will look smaller by a factor of $1-\betao$ in
the direction of the motion and bigger by $1+\betao$ in the opposite
direction. This implies that in the direction of the observers motion cluster
profiles will seem to be a little steeper and more concentrated.

\begin{figure*}[t]
\centering
\includegraphics[width=\plotwdtwo]
{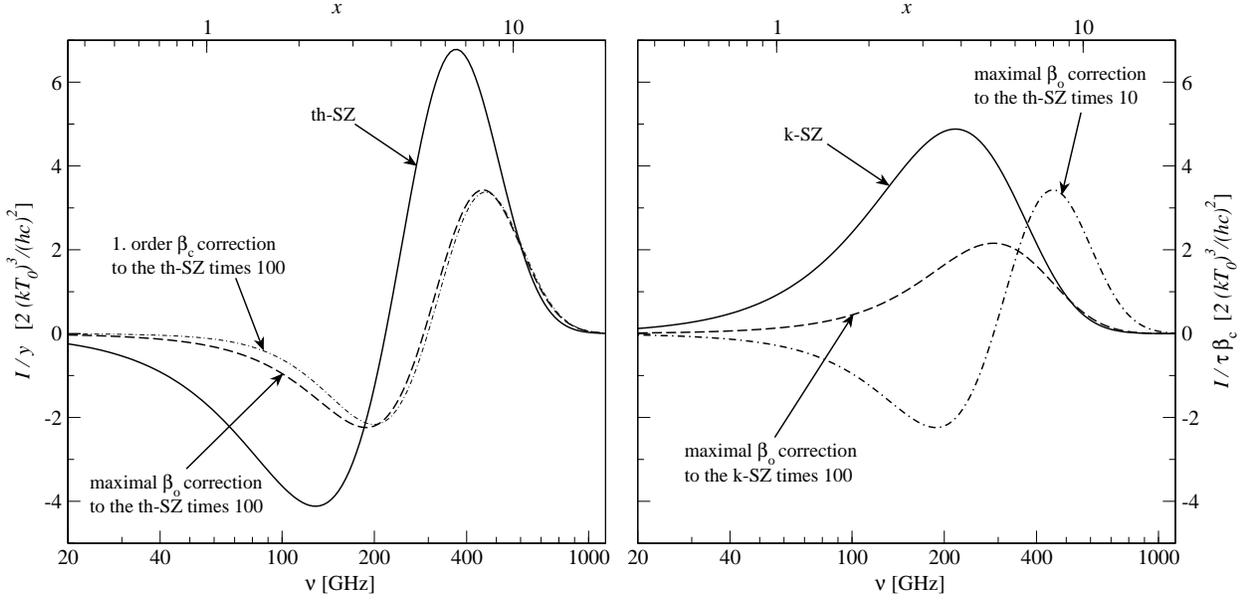}
\caption{Frequency-dependence of the SZ brightness due the non-relativistic
th- and k-SZ and the corresponding corrections induced by the motion of the
Solar System relative to the CMB rest frame for a cluster with electron
temperature $\kB \Te=5.1\,$keV, which is moving with $\betac=10^{-3}$ towards
the observer and is located at the maximum of the CMB dipole,
i.e. $\betaopar=1.241\cdot 10^{-3}$. In addition the first order $\betac$
correction to the th-SZ (dash-dotted line in the left panel) is shown. Note
that for convenience all the velocity corrections are multiplied by some factor
as given in the labels.}
\label{fig:YCD}
\end{figure*}

\subsection*{Transformation of the measured flux}
The spectral flux $F(\nu, \vek{n}_0)$ from a solid angle area $\mathcal{A}$ in
the direction $\vek{n}_0$ on the sky in $S$ is given by the integral
$F(\nu, \vek{n}_0)=\int\nolimits_\mathcal{A} I(\nu,\vek{n})\,\mu_0\id \Omega$,
where we defined $\mu_0=\vek{n}\cdot\vek{n}_0$.
If one assumes that the angular dimension of $\mathcal{A}$ is small, then using
\eqref{eq:transIapp} in the observers frame $S'$ the change of the flux due to
Doppler boosting and aberration is given by
\bsub
\label{eq:DFnuD}
\beal
\label{eq:DFnuDa}
\Delta F'_{\rm D}(\nu',\vek{n}'_0)
&\approx\betaopar\,\left[3-\nu'\partial_{\nu'}\right] F(\nu',\vek{n}'_0)
\\[1mm]
\Delta F'_{\rm A}(\nu',\vek{n}'_0)
&\approx\betaopar\,\int\nolimits_\mathcal{A}\,\theta'^2\,\partial_{\theta'} I(\nu',\vek{n}')\id \theta'\id\phi'
\Abst{.}
\end{align}
\esub
Assuming that the area $\mathcal{A}$ contains an unresolved object, which
contributes most of the total flux and vanishes on the boundaries of the
region, then the term arising due to aberration only can be rewritten as
\beal
\label{eq:DFnuA}
\Delta F'_{\rm A}(\nu',\vek{n}'_0)&=-2\,\betaopar\, F(\nu',\vek{n}'_0)
\Abst{.}
\end{align}
This can be simply understood considering that in the direction of the motion
the solid angle covered by an object is smaller by a factor
$[1-\betaopar]^2\approx1-2\,\betaopar$.
In this case the total change in the spectral flux $F'(\nu',\vek{n}'_0)$ is
\beal
\label{eq:DFnutot}
\frac{\Delta F}{F}=\frac{F'(\nu',\vek{n}'_0)-F(\nu',\vek{n}'_0)}{F(\nu',\vek{n}'_0)}
=\betaopar\left[1-\pAb{\ln F}{\ln\nu'}\right]
\Abst{.}
\end{align}
Integrating the flux $F'(\nu', \vek{n}'_0)$ over frequency $\nu'$ it is
straightforward to obtain the change of the total bolometric flux $F'_{\rm
bol}(\vek{n}'_0)=\int F'(\nu', \vek{n}'_0)\id\nu'$ in the observers frame
$S'$:
\beal
\label{equ:DFtot}
\frac{\Delta F_{\rm bol}}{F_{\rm bol}}=\frac{F'_{\rm bol}(\vek{n}'_0)-F_{\rm bol}(\vek{n}'_0)}{F_{\rm bol}(\vek{n}'_0)}
=2\,\betaopar
\Abst{.}
\end{align}
This results can also be easily understood considering the transformation law
for the total bolometric intensity $I_{\rm bol}=\int I(\nu)\id\nu$,
i.e. $I'_{\rm bol}=I_{\rm bol}/[\gamma(1-\betao\mu')]^4$, and the
transformation of the solid angle $\id
\Omega'=[\gamma(1-\betao\mu')]^2\id\Omega$.

\subsection*{Transformation of the number counts}
Defining $\frac{\id N}{\id\Omega}(F,\vec{n})$ as the number of objects per
solid angle $\id \Omega$ above a given flux $F$ at some fixed frequency $\nu$
and in some direction $\vec{n}$ on the sky in the CMB rest frame $S$, then the
corresponding quantity in the observers frame $S'$ is given by
\beal
\label{eq:dNdO}
\Abl{N'}{\Omega'}(F',\vec{n}')&=\Abl{N}{\Omega}(F,\vec{n})\,\Abl{\Omega}{\Omega'}
\Abst{,}
\end{align}
where $F$ and $\vek{n}$ are functions of $F'$ and $\vek{n}'$. Now, assuming
isotropy in $S$, in first order of $\betaopar$ one may write
\bmulti
\label{eq:dNdOp}
\Abl{N'}{\Omega'}(F',\vec{n}')
\approx\Abl{N}{\Omega}(F')
\\
\times \left[1+2\,\betaopar-\frac{\Delta F}{F}\,\frac{\partial\ln \frac{{\rm d}N}{{\rm d}\Omega}(F')}{\partial\ln F'}\right]
\Abst{,}
\end{multline}
with $\Delta F=F'-F$. 
For unresolved objects $\Delta F/F$ is given by equation
\eqref{eq:DFnutot}. Here we made use of the transformation law for the solid
angles and performed a series expansion of $\frac{\id N}{\id\Omega}(F)$ around
$F'$.

If one assumes $\frac{\id N}{\id\Omega}(F)\propto F^{-\lambda}$ and
$F(\nu)\propto \nu^{-\alpha}$, it is straightforward to show that for
unresolved sources $\frac{\id N'}{\id\Omega'}(F',\vec{n}')\approx\frac{\id
N}{\id\Omega}(F')\left[1+\betaopar(2+\lambda[1+\alpha])\right]$. This result
was obtained earlier by \citet{Ellis84} for the change of the radio source
number counts due to the motion of the observer. Dependent on the sign of the
quantity $\Sigma=2+\lambda[1+\alpha]$ there is an increase or decrease in the
number counts in one given direction. However, in the case of clusters
$\alpha$ is a strong function of frequency, which makes the situation more
complicated.

\section{Transformation of the cluster signal}
\label{sec:trans_cl}
For an observer at rest in the frame $S$ defined by the CMB the change of the
{\it surface brightness} in the direction $\vec{n}$ towards a cluster of
galaxies is given by the sum of the signals due to the th-SZ, $\DIth(\nu,
\vec{n})$ and the k-SZ, $\DIk(\nu, \vec{n})$:
\beal
\label{eq:DI_SZE_tot}
\DISZ(\nu, \vec{n})=\DIth(\nu, \vec{n})+\DIk(\nu, \vec{n})
\Abst{.}
\end{align}
In the non-relativistic case these contributions are given by
\citep[see][]{Zeld69, Sun80a, Sun80}:
\bsub
\label{eq:SZE}
\beal
\label{eq:SZEb}
\DIth(\nu, \vec{n})&=y
\,\frac{\xn\,e^{\xn}}{e^{\xn}-1}\left[\xn\,\frac{e^{\xn}+1}{e^{\xn}-1}-4\right]I_0(\nu)\\[1mm]
\label{eq:SZEc}
\DIk(\nu, \vec{n})&=\tau\,\betacpar\,\frac{\xn\,e^{\xn}}{e^{\xn}-1}\,I_0(\nu)
\Abst{,}
\end{align}
\esub
where $I_0(\nu)=\frac{2 h}{c^2}\,\frac{\nu^3}{e^\xn-1}$ denotes the CMB
monopole intensity with temperature $T_0$, $y=\int \frac{\kB\Te}{\me
c^2}\,\Ne\,\sigT\id l$ is the Compton $y$-parameter, with the electron number
density $\Ne$, and we introduced the abbreviation $x=h\nu/\kB T_0$.
Here we are only interested in the correction to the intensity in the central
region of the cluster, where the spatial derivative of $y$ is small and the
effects of aberration may be neglected. Using equations \eqref{eq:transIappa}
and \eqref{eq:SZE} one may find
\bsub
\label{eq:SZE_trans}
\beal
\label{eq:SZE_transb}
\DIthp(\nu',\vec{n}')&=y\,I_0(\nu')\,\frac{x' e^{x'}}{e^{x'}-1}\,\gth(x')\cdot \betao\,\mu'\\[1mm]
\label{eq:SZE_transc}
\DIkp(\nu',\vec{n}')&=\tau\,\betacpar\,I_0(\nu')\,\frac{x' e^{x'}}{e^{x'}-1}\,\gkin(x')\cdot \betao\,\mu'
\end{align}
\esub
for motion-induced change of the cluster brightness. Here the functions
$\gth(\x)$ and $\gkin(\x)$ are defined by
\bsub
\label{eq:g_X}
\beal
\gth(\x)&=4-6\,\mathcal{X}+\mathcal{X}^2+\frac{1}{2}\mathcal{S}^2\\
\gkin(\x)&=\mathcal{X}-1
\end{align}
\esub
and the notations $\mathcal{X}=x\,\coth(\frac{x}{2})$,
$\mathcal{S}=x/\sinh(\frac{x}{2})$ were introduced.

In Fig. \ref{fig:YCD} the spectral dependence of $\DIthp$ and $\DIkp$ is
illustrated. The transformation of the th-SZ leads to a spectral distortion
which is very similar to the first order $\betac$ correction to the th-SZ. In
the Rayleigh-Jeans limit $\gth(\x)\rightarrow -2$ and therefore is 5 times
bigger than the $\betac$ correction to the th-SZ. The maximum/minimum of
$\DIthp$ is at $x=7.97\,/\,3.31$ and $\DIthp$ vanishes at $x=5.10$ ($x=1$
corresponds to $\nu=56.8\,$GHz).
On the other hand the transformation of the k-SZ leads to a y-type spectral
distortion with the corresponding y-parameter
$\ykinp=\tau\,\betacpar\,\betao\,\mu'\sim 10^{-8}\,\mu'$. 
The maximum of $\DIkp$ is at $x=5.10$. Fig. \ref{fig:YCD} clearly shows, that
the motion-induced correction to the th-SZ easily reaches the level of a few
percent in comparison to the k-SZ (e.g. at $\nu=400\,$GHz it contributes $\sim
14$\% to the k-SZ signal for a cluster with $\kB\Te=5.1\,$keV and
$\betac=10^{-3}$).
\begin{figure*}
\centering
\includegraphics[width=\plotwdtwo]
{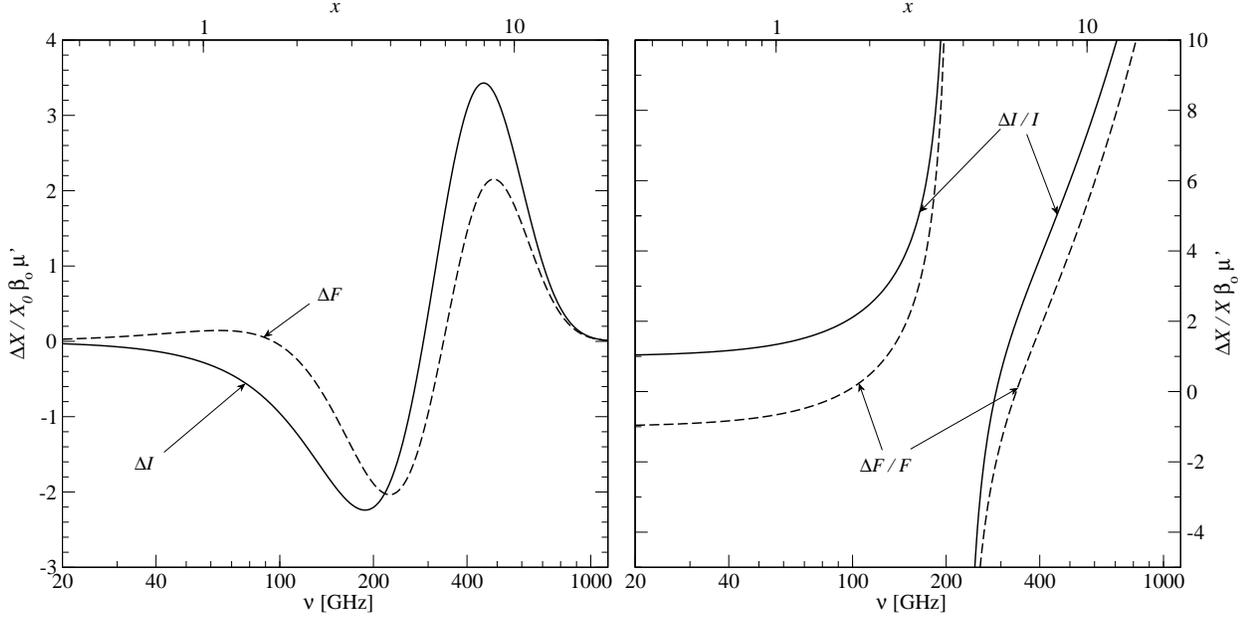}
\caption{Motion-induced change of the CMB spectral brightness $\Delta I$ and
flux $\Delta F$ for a cluster of galaxies resting with respect to the
CMB. {\bf Left panel}: Absolute change of the cluster brightness (solid line)
with $X_0=2.70\times 10^{11}\,y\,$mJy/sr and the flux for an unresolved
cluster (dashed line) with $X_0=7.17\,\left<y/10^{-4}\,r_{\rm c}^2\right>_{\rm
cl}(r_{\rm c}/30'')^2\,$mJy,
where $r_{\rm c}$ is the core radius and where $\left<u\right>_{\rm
cl}=\int_{\rm cl} u\id\Omega/4\pi$ denotes the cluster average of the quantity
$u$.
{\bf Right panel}: Relative change of the brightness (solid line) and flux for
an unresolved cluster (dashed line). Here $\Delta I/I=\betao\,\mu'\,
(3+\alpha)$ and $\Delta F/F=\betao\,\mu'\,(1+\alpha)$ (cf.
Equ. \eqref{eq:IFdN}).}
\label{fig:DIDF}
\end{figure*}

In order to obtain the motion-induced change of the {\it flux} for unresolved
clusters one has to integrate the surface brightness over the surface of the
cluster. In the following we neglect the k-SZ, since its contribution only
becomes important close to the crossover frequency. Then it follows that
$\Delta F(\nu, \vec{n})\propto \Delta I(\nu, \vec{n})$, implying that $\id\ln
F/\id\ln \nu'=\id\ln I/\id\ln x'$. Comparing equations \eqref{eq:transIappa}
and \eqref{eq:SZE_trans} one can define the effective spectral index of the SZ
signal by
\beal
\label{eq:dlnI_dx'}
\alpha=-\frac{\id\ln F(x')}{\id\ln x'}
=\frac{\gth(x')}{Y_0(x')}-3
\Abst{,}
\end{align}
with $Y_0(x)=\mathcal{X}-4$. 
Using equation \eqref{eq:dlnI_dx'} one can write the central {\it brightness},
{\it flux} and {\it number count} for unresolved clusters as
\bsub
\label{eq:IFdN}
\beal
\DISZp(\nu', \vec{n}')&=\DISZ(\nu', \vec{n}')\,\left[1+\betao\,\mu'(3+\alpha)\right]
\\[1mm]
\DFSZp(\nu', \vec{n}')&=\DFSZ(\nu', \vec{n}')\,\left[1+\betao\,\mu'(1+\alpha)\right]
\\[1mm]
\label{eq:IFdNc}
\frac{\id N'_{\rm SZ}}{\id\Omega'}(F',\vec{n}')
&=\frac{\id N_{\rm SZ}}{\id\Omega}(F')\,\left[1+\betao\,\mu'\Sigma\right]
\Abst{,}
\end{align}
\esub
with $\Sigma=2+\lambda(1+\alpha)$ and $\lambda=-\frac{\partial \ln \frac{\id
N}{\id\Omega}(F')}{\partial \ln F'}$.
Fig. \ref{fig:DIDF} shows the change of the central brightness and the flux
for an unresolved cluster. It is obvious that only in the RJ region of the CMB
spectrum the SZ brightness and flux follow a power-law. The change of the
number counts will be discussed below (see Sect. \ref{sec:number}).

\section{Multi-frequency observations of clusters}
\label{sec:Observations}
The observed CMB signal in the direction of a cluster consists of the sum of
all the contributions mentioned above, including the relativistic correction
to the SZ effect. Given a sufficient frequency coverage and spectral
sensitivity one may in principle model the full signal for even one single
cluster, but obviously there will be degeneracies which have to be treated
especially if noise and foregrounds are involved.
Therefore it is important to make use of the special properties of each
contribution to the total signal, such as their spectral features and spatial
dependencies.

One obstacle for any multi-frequency observation of cluster is the
cross-calibration of different frequency channels. Some method to solve this
problem was discussed in \citet{Chluba2004} using the spectral distortions
induced by the superposition of blackbodies with different temperatures. In
the following we assume that the achieved level of cross-calibration is
sufficient.

The largest CMB signal in the direction of a cluster (after elimination of the
CMB dipole) is due to the th-SZ. In order to handle this signal one can make
use of the zeros of the spectral functions describing the relativistic
corrections. In addition, future X-ray spectroscopy will allow us to accurately
determine the mean temperature of the electrons inside clusters. This
additional information will place useful constraints on the parameters
describing the th-SZ and therefore may bring us down to the effects connected
with the peculiar velocities of the cluster and the observer.

The temperature difference connected with the non-relativistic k-SZ is
frequency-independent and therefore may be eliminated by multi-frequency
observations. As mentioned above (see Fig. \ref{fig:YCD}) the motion-induced
spectral distortion to the th-SZ has an amplitude and spectral dependence,
which is very similar to the effect connected with the first order $\betac$
correction to the th-SZ. 
For many clusters on the other hand one can expect that the signals
proportional to $\betacpar$ average out. This implies that for large cluster
samples ($\sim 10^3-10^4$) only the signals connected with the th-SZ are
important.


\subsection{Dipolar asymmetry in the number of observed clusters}
\label{sec:number}
Integrating \eqref{eq:IFdNc} over solid angle leads to the observed number of
clusters in a given region of the sky. If one assumes that the observed region
is circular with radius $\thrp$ centered in the direction
$\vek{n}_0'=(\phi'_0, \theta_0')$ then the total observed number of clusters
is given by
\beal
\label{eq:N}
N'_{\rm SZ}(F')&=N_{\rm SZ, eff}(F')\,\left[1+\betao\,\mu'_0\,\frac{1+\murp}{2}\,\Sigma\right]
\Abst{,}
\end{align}
where $N_{\rm SZ, eff}(F')=4\pi\,\frac{\id N_{\rm
SZ}}{\id\Omega}\,\frac{1-\murp}{2}$ is the effective number of clusters inside
the observed patch with fluxes above $F'$, $\murp=\cos \thrp$ and $\mu_0'=\cos
\theta_0'$.
For two equally sized patches in separate directions on the sky the difference
in the number of observed clusters will then be
\beal
\label{eq:DN}
\Delta N'=\betao\,N_{\rm SZ, eff}\,\Delta\mu'_0\,\frac{1+\murp}{2}\,\Sigma
\Abst{,}
\end{align}
with $\Delta\mu'_0=\mu'_{0,1}-\mu'_{0,2}$, where $\mu'_{0,i}=\cos
\theta_{0,i}'$ for patch $i$. Centering the first patch on the maximum and the
second on the minimum of the CMB dipole leads to the maximal change in the
number of observed clusters at a given frequency ($\Delta\mu'_0=2$).
To estimate the significance of this difference we compare $\Delta N'$ to the
Poissonian noise in the number of clusters for both patches, which is given by
$\sqrt{N'_{\rm SZ,1}+N'_{\rm SZ,2}}\approx\sqrt{2\,N_{\rm SZ, eff}}$. To
obtain a certain signal to noise level $q$ the inequality
\beal
\label{eq:sig}
\betao\,|\Sigma|\,\Delta\mu'_0\,[1+\murp]\,\sqrt{1-\murp} 
\geq \frac{4\,q}{\sqrt{N_{\rm SZ, f}}}
\end{align}
has to be fulfilled. We defined $N_{\rm SZ, f}=4\pi\,\frac{\id N_{\rm
SZ}}{\id\Omega}$ as the number of clusters on the whole sky above a given flux
level $F'$. Here two effects are competing: the smaller the radius of each
patch, the smaller the number of observed clusters above a given flux but the
larger the effective $\left<\betaopar\right>$. The optimal radius is
$\thr\sim~70^\circ$ but for a given $q$ and sensitivity the size in
principle can be smaller.


\subsection{Numerical estimates for the dipolar asymmetry in the cluster number counts}
\begin{figure}
\centering \includegraphics[width=\plotwd] {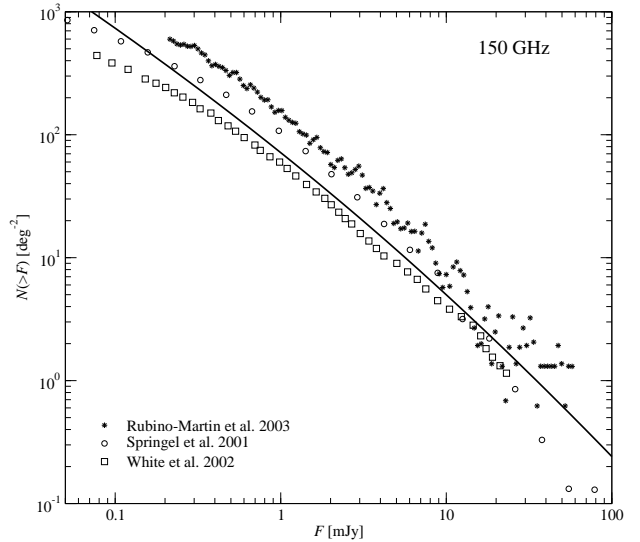}
\caption{$\log N$-$\log F$: Number of clusters per square degree with flux
level above $F$ at observing frequency $\nu=150\,$GHz.  The solid line shows
the modified Press-Schechter prescription as used in this work.}
\label{fig:logNlogS}
\end{figure}
In this section we present results for the SZ cluster number counts using a
simple Press-Schechter \citep{PS1974} prescription for the mass function of
halos as modified by \citet{SMT2001} to include the effects of ellipsoidal
collapse.
Since we are interested in unresolved objects we only need to specify the
cluster mass-temperature relation and baryonic fraction. 
For the former we apply the frequently used scaling relation and normalization
as given by \citet{BN1998}, whereas for the latter we simply assume a
universal value of $\Omega_{\rm b}/\Omega_{\rm m}$, which is rather close to
the local values as derived from X-ray data \citep[e.g.][]{MME1999} independent of
cluster mass and redshift. We note that these two assumptions are the biggest
source of uncertainty in our calculations and the use of them is only
justified given the lack of current knowledge about the detailed evolution of
the baryonic component in the Universe.
In spite of these gross simplifications our results on cluster number counts
agree very well with those obtained in state-of-the-art hydrodynamical
simulations by \citet{SWH2001} and \citet{WHS2002} as demonstrated in
Fig. \ref{fig:logNlogS}. Here the counts are calculated for the $\Lambda$CDM
concordance model \citep{Spergel2003} assuming observing frequency of
$150\,$GHz. The first set of simulations included only adiabatic gas physics
whereas for the second also gas cooling and feedback from supernovae and
galactic winds was taken into account. We also plot the results obtained by
\citet{Alberto2003} using a Monte-Carlo simulations based on a Press-Schechter
approach. In the estimates presented below we will use the curve given by the
solid line in Fig.\ref{fig:logNlogS}, which in the most interesting range of
lower flux limits ($1\,$mJy -$10\,$mJy) has an effective power-law slope of
$\lambda\sim 1.10-1.25$.

\begin{figure}
\centering
\includegraphics[angle=270, width=\plotwd]
{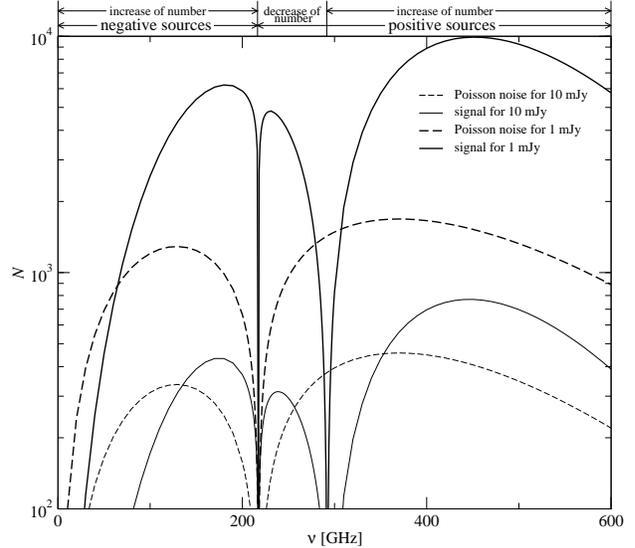}
\caption{Motion-induced dipolar asymmetry in number counts (solid lines) as a
function of the observing frequency using the optimal patch radius
$\theta_{\rm r}=70^\circ$ for both patches, where the first is centered on the
maximum, the second on the minimum of the CMB dipole. For comparison we give
the corresponding $1\sigma$ Poissonian noise level (dashed curves).
In addition we mark the regions where we expect an increase of the number of
negative sources and a decrease/increase of the number of positive sources,
respectively, if one is observing only in the direction of the maximum of the
CMB dipole.}
\label{fig:numb}
\end{figure}

In Fig. \ref{fig:numb} we compare the motion-induced dipolar asymmetry in
number counts as a function of the observing frequency using the optimal patch
radius $\theta_{\rm r}=70^\circ$ for both sides of the sky with the $1\sigma$
Poisson noise level for the two lower flux limits of $1\,$mJy and $10\,$mJy.
In addition we mark the regions where we expect an increase of the number of
negative sources and a decrease/increase of the number of positive sources,
respectively, if one is observing only in the direction of the maximum of the
CMB dipole.
It is important to note that the motion-induced change in the cluster number
counts vanishes at frequencies $\nu\sim~300\,$GHz. The exact value of this
frequency depends both on the slope of the number count curve and the spectral
index.

\begin{figure}
\centering
\includegraphics[width=\plotwd]
{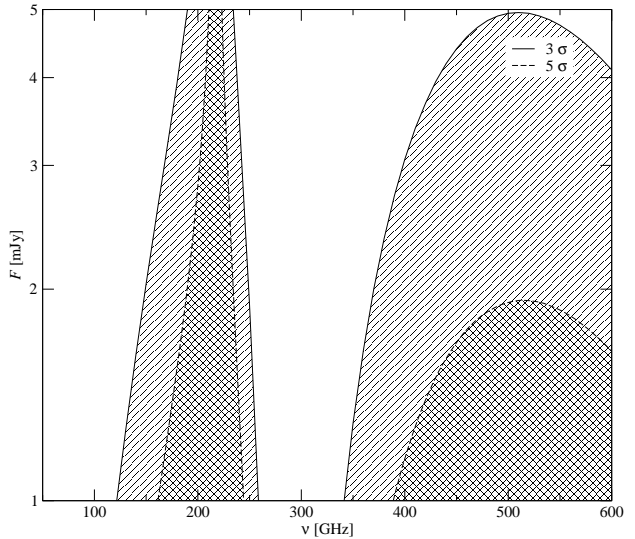}
\caption{Required sensitivities for a $3\sigma$ and $5\sigma$-detection of the
motion-induced dipolar asymmetry in number counts as a function of
frequency. The shaded areas indicate regions where a detection above
$3\,\sigma$ and $5\,\sigma$ level is possible, respectively.}
\label{fig:limits}
\end{figure}

\begin{figure}
\centering
\includegraphics[width=\plotwd]
{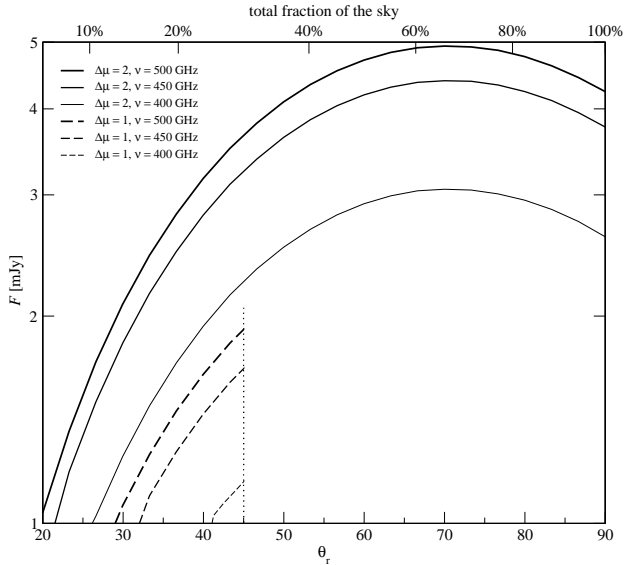}
\caption{Required sensitivity limits for a $3\sigma$ detection of the number
count asymmetry in the frequency range $400-500\,$GHz as a function of patch
radius. One patch is centered on the maximum of the CMB dipole. Note that for
$\Delta \mu=1$ and $\theta_{\rm r}>~45^\circ$ the two patches start to overlap
and therefore the corresponding curves were not presented here.}
\label{fig:limits_mju}
\end{figure}

Fig. \ref{fig:limits} presents the sensitivity limits where the motion-induced
signal is equal to the $3\sigma$ and $5\sigma$ Poissonian noise levels for
different observing frequencies.
Taking into account that new generation of SZ dedicated surveys (e.g. {\sc
Spt} and {\sc Act}) will have mJy sensitivities, we see that for experiments
covering the full sky a detection of the motion-induced signal is clearly
within the reach of the capabilities of modern technology especially at
frequencies in the range $\nu=400-500\,$GHz. Combining the data of different
experiments with limited sky coverage may lead to a sufficient total sample
(see Fig. \ref{fig:limits_mju}).

It is important to mention that in our simplistic calculations we assumed that
all the clusters remain unresolved, which is a good approximation for the {\sc
Planck}, {\sc Spt} and {\sc Act}. For experiments with higher angular
resolution a significant population of clusters will be resolved and hence the
number count curves presented here will change accordingly. 

From Fig. \ref{fig:limits} we also see that the most promising frequencies for
a detection of the motion-induced asymmetries are around the crossover
frequency (i.e.$\sim 217\,$GHz) and in the range $\nu\sim
400-500\,$GHz. Clearly, for a proper modeling near the crossover frequency one
has to take into account the contribution from the k-SZ, which has been
neglected so far. It is evident that the k-SZ is contributing symmetrically to
channels around $217\,$GHz in the sense that the number of positive and
negative sources is approximately equal. On the other hand in the range
$\nu\sim 400-500\,$GHz other astrophysical source start to contribute to the
source counts (see Sect. \ref{sec:dustG}).

Finally, in Fig. \ref{fig:limits_mju} we illustrate the dependence of the
required sensitivities for a $3\sigma$ detection of the number count asymmetry
on the radius of the two compared patches close to the frequency $\nu\sim
500\,$GHz. The first patch is centered on the maximum of the CMB dipole,
whereas for the second we choose the two cases $\theta_{0,2}=90^\circ$ and
$180^\circ$, i.e. $\Delta \mu_0=1$ and $2$, respectively.
This Figure shows that for frequencies below $\sim 400-500\,$GHz and
separation angles smaller than $90^\circ$ a detection of the asymmetry will
only be feasible for experiments with sub-mJy sensitivity.

\subsection{Source count contribution from non-SZ populations}
\label{sec:dustG}
In the range $\nu\sim 400-500\,$GHz, which is most promising for a detection
of the motion-induced number count asymmetry, other foreground sources begin
to play a role, e.g. dusty high redshift galaxies \citep{Blain2002}. In the
microwave band these galaxies have extremely peculiar spectrum $F(\nu)\propto
\nu^{-\alpha_{\rm d}}$, with $\alpha_{\rm d}$ ranging from $-3$ to $-4$.
Using formula \eqref{eq:dNdOp} it is easy to show that the observed properties
of this population will also be influenced by the motion of the Solar System,
but in a completely different way than clusters: in the direction of our
motion relative to the CMB rest frame their brightness and fluxes decrease
when for clusters they increase. This implies that in the frequency range
$\nu\sim 400-500\,$GHz the motion-induced dipolar asymmetry in the number
counts for these sources has the opposite sign in comparison to clusters,
i.e. $\Delta N'_{\rm d}<0$ when $\Delta N'_{\rm cl}>0$. 
Detailed multi-frequency observations should allow distinguishing the source
count contributions of these two classes of objects, but nevertheless it is
interesting that they a different sign of the motion-induced flux dipole.

\section{Conclusion}
In this paper we derived the changes to the SZ cluster brightness, flux and
number counts induced by the motion of the Solar System with respect to the
CMB rest frame. These corrections to the SZ cluster brightness and flux have
similar spectral behavior and amplitude as the first order velocity correction
to the th-SZ (see Fig. \ref{fig:YCD}) and thus need to be taken into account
for the precise modeling of the cluster signal. Since both the amplitude and
direction of the motion of the Solar System is known with a high precision it
is easy to correct for these changes.

The dipolar asymmetry induced in the SZ cluster number counts in contrast to
the counts of more conventional sources can change polarity dependent on the
observational frequency (see Sect. \ref{sec:dustG}). This behavior is due to
the very specific frequency dependence of the SZ effect. 
We find that frequencies around the crossover frequency $\sim 217$ GHz and in
the range $\sim 400-500$ GHz are the most promising for a detection of this
motion-induced number count asymmetry (see Fig. \ref{fig:limits}).

\acknowledgements{G.H. acknowledges the support provided through the European
Community's Human Potential Programme under contract HPRN-CT-2002-00124,
CMBNET, and the ESF grant 5347.}

\end{document}